\documentclass[aps,prd,twocolumn,preprintnumbers,nofootinbib,amsmath,amssymb]{revtex4-2}

\usepackage[normalem]{ulem}
\usepackage{xcolor}

\usepackage[pdftex, colorlinks=true, linkcolor=black, citecolor=black,
urlcolor=black,pdftitle={Comment on ``Chiral symmetry restoration, the
  eigenvalue density of the Dirac operator, and the axial U(1) anomaly
  at finite temperature''}]{hyperref}

\newcommand{\f}[2]{\frac{#1}{#2}}

\newcommand{\la}{\langle}
\newcommand{\ra}{\rangle}

\newcommand{\Oc}{\mathcal{O}}
\newcommand{\Lc}{\mathcal{L}}

\newcommand{\chit}{\chi_t}

\makeatletter
\newcommand{\superimpose}[3][\mathord]{#1{\mathpalette\superimpose@{{#2}{#3}}}}
\newcommand{\superimpose@}[2]{\superimpose@@{#1}#2}
\newcommand{\superimpose@@}[3]{%
  \ooalign{%
    \hfil$\m@th#1#2$\hfil\cr
    \hfil$\m@th#1#3$\hfil\cr
  }%
}
\makeatother

\newcommand{\Nc}{\superimpose{\backslash}{\mathcal{L}}}

\newcommand{\Pc}{\mathcal{P}}
\newcommand{\Sc}{\mathcal{S}}

\newcommand{\dl}{\delta}

\newcommand{\eqdef}{\mathrel{\mathop:}=}

\begin{document}

\title{Comment on ``Chiral symmetry restoration, the eigenvalue
  density of the Dirac operator, and the axial U(1) anomaly at finite
  temperature''}

\author{Matteo Giordano} \email{giordano@bodri.elte.hu}
\affiliation{Institute of Physics and Astronomy, ELTE E\"otv\"os
  Lor\'and University, P\'azm\'any P\'eter s\'et\'any 1/A, H-1117,
  Budapest, Hungary}

\date{\today}

\begin{abstract}
  Aoki, Fukaya, and Taniguchi claim that both the spectral density of
  the Dirac operator at the origin and the topological susceptibility
  must vanish identically for sufficiently small but nonzero quark
  mass $m$ in the chirally symmetric phase of quantum chromodynamics
  with two light quark flavors, under certain technical assumptions on
  the spectrum and on the dependence of observables on $m$. I argue
  that a crucial step of their proof is not justified, and the
  validity of these conclusions should be reassessed.
\end{abstract}

\maketitle

\paragraph{}

In Ref.~\cite{Aoki:2012yj} the authors study the constraints imposed
by the restoration of $N_f=2$ chiral symmetry on the spectrum of the
Dirac operator in finite-temperature quantum chromodynamics on the
lattice. Their conclusions, based on certain technical assumptions on
the spectral density and on the mass dependence of certain
observables, include the claims that (i.)\ the spectral density of
Dirac eigenvalues, $\rho(\lambda;m)$, vanishes identically at
$\lambda=0$ for sufficiently small quark mass, $m$, i.e.,
$\rho(0;m)= 0$ $\forall |m|<m_0$, for some $m_0>0$; and (ii.)\ the
topological susceptibility, $\chit$, vanishes identically for
sufficiently small $m$. These conclusions have important consequences
for the fate of the anomalous $\mathrm{U}(1)_A$ symmetry in the chiral
limit, and are used in Ref.~\cite{Aoki:2012yj} to argue that
$\mathrm{U}(1)_A$ is effectively restored in the chiral limit in the
scalar and pseudoscalar sector as a necessary consequence of chiral
symmetry restoration. Notably, (ii.)\ would indicate the presence of a
phase transition at small but nonzero $m$ as one goes toward the
chiral limit in the high-temperature phase.

Predictions (i.)\ and (ii.)\ have been extensively tested with
numerical lattice simulations toward the chiral
limit~\cite{Cossu:2013uua,Tomiya:2016jwr,Aoki:2020noz,JLQCD:2024xey,
  Chandrasekharan:1998yx,Dick:2015twa,HotQCD:2019xnw,Ding:2020xlj,
  Kaczmarek:2021ser}. However, while Refs.~\cite{Cossu:2013uua,
  Tomiya:2016jwr,Aoki:2020noz,JLQCD:2024xey} support these
predictions, Refs.~\cite{Chandrasekharan:1998yx,Dick:2015twa,
  HotQCD:2019xnw,Ding:2020xlj,Kaczmarek:2021ser} reach different
conclusions. On the one hand, Refs.~\cite{Cossu:2013uua,
  Tomiya:2016jwr,Aoki:2020noz,JLQCD:2024xey} secure a better control
of theoretical uncertainties by using M\"obius domain-wall fermions,
at the price of computational limitations on the available volumes and
number of configurations. On the other hand,
Refs.~\cite{Chandrasekharan:1998yx,Dick:2015twa,HotQCD:2019xnw,
  Ding:2020xlj,Kaczmarek:2021ser} gain access to larger volumes and
statistics by using staggered fermions, but introduce theoretical
uncertainties on the chiral and topological properties of the system
by doing so. Owing to these difficulties no consensus has been reached
yet, and the status of these predictions remains unsettled.

\paragraph{}

The proof of (i.)\ and (ii.)\ given in Ref.~\cite{Aoki:2012yj} makes
crucial use of the following argument. In Sec.~II~C of
Ref.~\cite{Aoki:2012yj} [see their Eqs.~(23) and (24)], the authors
state that if for some mass-independent positive-definite observable
$\Oc(A)$, involving only (lattice) gauge fields, $A$, one has for some
integers $l_0\ge 1$ and $k\ge 0$ that
\begin{equation}
  \label{eq:aft_crit4_0}
  \lim_{m\to 0} \f{1}{m^k} \la \Oc(A)^{l_0}\ra = 0\,,
\end{equation}
where $\la\ldots\ra$ denotes the expectation value, then one can write
\begin{equation}
  \label{eq:aft_crit4}
  \la \Oc(A)^{l_0}\ra =   m^{k_0}\int D\!A \, P(m,A)\Oc(A)^{l_0}\,,  
\end{equation}
with $k_0>k$, where $D\!A$ is the gauge-field integration measure and
$P(m,A)$ a non-negative quantity that does not vanish identically in
the chiral limit, $P(0,A)\not\equiv 0$. It is assumed that the
integral in Eq.~\eqref{eq:aft_crit4} is finite (i.e., not divergent)
in the infinite-volume limit and, implicitly, also in the subsequent
chiral limit; it is also implicitly assumed that $P(0,A)$ is well
defined. According to the authors, this means that the leading $m$
dependence of $\la \Oc(A)^{l_0}\ra$ arises from those gauge
configurations for which $P(0,A)\neq 0$. From Eq.~\eqref{eq:aft_crit4}
they then infer that
\begin{equation}
  \label{eq:aft_crit5}
  \la \Oc(A)^{l}\ra = m^{k_0} \int D\!A \, P(m,A) \Oc(A)^{l} = O(m^{k_0})\,,  
\end{equation}
for arbitrary positive integer $l$ [see their Eq.~(25)], as long as
the integral is finite (in the infinite-volume limit,
presumably). According to the authors, this is because $\Oc(A)^{l_0}$
and $\Oc(A)^{l}$ are both positive and share the same support in
configuration space.

\paragraph{}

It is inherent in the nature of the problem that the expectation value
$\la\ldots\ra$ in Eqs.~\eqref{eq:aft_crit4_0}--\eqref{eq:aft_crit5}
must be understood as the thermodynamic limit of the expectation value
computed in a finite volume; this meaning of the notation will be
understood in the following. In fact, for Eq.~\eqref{eq:aft_crit4_0}
to be a meaningful statement about the chiral limit, this should be
taken only after the thermodynamic limit. Similarly, the measure
$ D\!A \, P(m,A)$ appearing on the right-hand side of
Eqs.~\eqref{eq:aft_crit4} and \eqref{eq:aft_crit5} should be
understood as a measure formally defined on the space of
configurations of infinite volume.
 
By referring to Eq.~\eqref{eq:aft_crit4} as providing the leading $m$
dependence of $ \la \Oc(A)^{l_0}\ra$ [see after their Eq.~(24)], the
authors of Ref.~\cite{Aoki:2012yj} are implicitly assuming that the
integral does neither diverge nor vanish in the chiral limit, and so
not only that $ P(0,A)$ is well defined and not identically zero, but
also that the measure of the support of $\Oc(A)^{l_0}$ according to
$D\!A \, P(0,A)$ is nonzero. Nonetheless, if this measure were zero
one would simply find that $\la \Oc(A)^{l_0}\ra$ vanishes faster than
$m^{k_0}$, and their argument would not change substantially.

Generally, Eq.~\eqref{eq:aft_crit4_0} does not imply
$\la \Oc(A)^{l_0}\ra=O(m^{k_0})$ with $k_0>k$, since
$m^{-k} \la \Oc(A)^{l_0}\ra$ may be vanishing more slowly than a
power. However, in Ref.~\cite{Aoki:2012yj} this is justified by the
assumption that in the symmetric phase the expectation value in the
thermodynamic limit (if it exists at $m\neq 0$) of a mass-independent
observable involving only gauge fields is an analytic function of
$m^2$ in a neighborhood of $m=0$ (referred to as ``$m^2$-analyticity
assumption'' in the following). This also allows the authors to write
$k_0= 2\left(\lfloor k/2\rfloor +1\right)$ and
$P(m,A)=\hat{P}(m^2,A)$, which is, however, immaterial for what
follows.  Since the $m^2$-analyticity assumption has far-reaching and
unexpected consequences, it is better to proceed at first without
imposing it. One should then understand a factor $m^{k_0}$ as denoting
the leading, $o(m^k)$ dependence of $ \la \Oc(A)^{l_0}\ra$ (changing
the notation is straightforward but would add to the clutter, without
any change in the arguments).

Finally, since the authors of Ref.~\cite{Aoki:2012yj} apply
Eq.~\eqref{eq:aft_crit5} to the topological charge squared divided by
the volume (see their Sec.~III~G), when they require that $\Oc(A)$ be
positive they most likely mean $\Oc(A)\ge 0$.  Here I will understand
positivity of an observable in the same way.

\paragraph{}

The authors of Ref.~\cite{Aoki:2012yj} do not provide details on how
they obtain Eq.~\eqref{eq:aft_crit4} from Eq.~\eqref{eq:aft_crit4_0},
and Eq.~\eqref{eq:aft_crit5} from Eq.~\eqref{eq:aft_crit4}. As one can
always write an $O(m^{k_0})$ expectation value in the form of
Eq.~\eqref{eq:aft_crit4} simply by factoring $m^{k_0}$ out from the
usual path integral, this equation is empty unless one requires that
$P(0,A)$ is well defined; and to show that it is so, one needs to
characterize $P(m,A)$ in more detail. Indeed, just factoring out
$m^{k_0}$ from the usual path-integral measure would generally lead to
a divergent $P(m,A)$ in the chiral limit. This can be seen explicitly
for a one-dimensional configuration space with measure
$dA\,\theta(m^2-|A|)/(2m^2)$, with $\theta(x)$ the Heaviside step
function, taking $\Oc(A) = |A|$ as the observable. In this example,
after observing that $\la \Oc(A)^{l_0}\ra=O(m^{k_0})$ with $k_0=2l_0$
one would tentatively set $P(m,A)=\theta(m^2-|A|)/(2m^{2(l_0+1)})$.
However, in the chiral limit $P(0,A)$ would not be well defined, not
even as a distribution on the space of polynomials of $\Oc(A)$ that
vanish at zero, and one would not be able to conclude that
$\la \Oc(A)^l\ra=O(m^{k_0})$, $\forall l>0$ (except if the initial
observation of a vanishing expectation value in the chiral limit had
been made for $l_0=1$). In fact, $\la \Oc(A)^l\ra=O(m^{2l})$, which is
not $O(m^{k_0})$ for $l<l_0$.

Moreover, $\Oc(A)^{l_0}$ and $\Oc(A)^{l}$ sharing the same support in
configuration space is by itself not sufficient to obtain
Eq.~\eqref{eq:aft_crit5} from Eq.~\eqref{eq:aft_crit4}, as in general
the configurations in this support that are relevant respectively to
$\la\Oc(A)^{l_0}\ra$ and $\la\Oc(A)^{l}\ra$ for $l\neq l_0$ have no
reason to provide contributions of the same order in $m$. The
$l$ independence of how fast $\la\Oc(A)^{l}\ra$ vanishes, if it does
so for some $l=l_0$, is an interesting technical claim on its own, so
it is worth discussing this point in detail.

A characterization of $P(m,A)$ that I believe is in the spirit of
Ref.~\cite{Aoki:2012yj}, and that allows one to derive
Eq.~\eqref{eq:aft_crit5} from Eq.~\eqref{eq:aft_crit4}, is the
following.  The arguments are heuristic, and any issue of rigor in the
construction of the various measures involved is ignored.  One starts
by writing the thermodynamic limit of the expectation value of a
generic observable $\mathcal{G}(A)$ depending only on gauge fields as
$\la\mathcal{G}(A)\ra = \int D\!A \, \mathcal{P}(m,A) \mathcal{G}(A)$,
for a suitable measure $D\!A \, \mathcal{P}(m,A)$ on the space of
configurations of infinite volume, assumed to be well behaved in the
chiral limit. This measure typically vanishes in most of configuration
space, and has support only on the configurations that remain relevant
in the thermodynamic limit. If one now assumes that the
(mass-dependent) support $S_m(\mathcal{P})$ of $\mathcal{P}(m,A)$ and
the (mass-independent) support $S(\Oc)$ of $\Oc(A)^{l_0}$, which is
the same as the support of $\Oc(A)$ (for $l_0\ge 1$), overlap in a
region of measure $O(m^{k_0})$ according to $D\!A\,\mathcal{P}(m,A)$,
and that in the chiral limit $\Oc(A)$ remains bounded in this region,
it is possible to define in a meaningful way the measure
$D\!A \,P(m,A)$ appearing in Eq.~\eqref{eq:aft_crit4}, leading in turn
to Eq.~\eqref{eq:aft_crit5}.  Indeed, denoting with $\chi_R(A)$ the
indicator function of the set of configurations $R$, and setting
$P_{\Oc}(m,A) \equiv m^{-k_0}\mathcal{P}(m,A) \chi_{S(\Oc)}(A)$, one
has 
\begin{equation}
  \label{eq:meas_shr}
  \la \Oc(A)^l \ra = m^{k_0} \int D\!A \, P_{\Oc}(m,A) \Oc(A)^{l}=O(m^{k_0})\,,  
\end{equation}
for any $l\ge 1$ (this construction of course does not apply to
$l=0$). In fact, since the effective support of $\Oc(A)$, i.e., the
support $S_m(\mathcal{P})\cap S(\Oc)$ of
$\chi_{S_m(\mathcal{P})}(A)\chi_{S(\Oc)}(A)$, is assumed to be of
measure $O(m^{k_0})$ according to $D\!A\,\mathcal{P}(m,A)$, after
rescaling by $m^{k_0}$ one obtains for the shrinking region of overlap
at most a finite nonzero measure according to $D\!A\, P_{\Oc}(m,A)$;
since $\Oc(A)$ is assumed to be bounded there, the integral is finite.

Summarizing, if the effective support $S_m(\mathcal{P})\cap S(\Oc)$ of
$\Oc(A)$ is of measure $O(m^{k_0})$ according to
$D\!A\,\mathcal{P}(m,A)$, then Eqs.~\eqref{eq:aft_crit4} and
\eqref{eq:aft_crit5} hold with the same meaningful $P(m,A)$ [i.e.,
$P_{\Oc}(m,A)$ defined above], nondivergent in the chiral limit. If
$\Oc(A)$ is bounded (at least on its effective support) this region
gives at most $O(1)$ contributions to the integrals in
Eqs.~\eqref{eq:aft_crit4} and \eqref{eq:aft_crit5}, that cannot weaken
the $O(m^{k_0})$ suppression given by the prefactor, independently of
$l$, and $ \la \Oc(A)^l \ra = O(m^{k_0})$ follows, $\forall l\ge 1$.

\paragraph{}

Even though Eq.~\eqref{eq:aft_crit5} is a necessary consequence of
Eq.~\eqref{eq:aft_crit4} if one accepts the assumptions and the
heuristic argument given above, Eq.~\eqref{eq:aft_crit4} with $P(m,A)$
having the required properties is not a necessary consequence of
Eq.~\eqref{eq:aft_crit4_0}. In fact, the relation
$\la \Oc(A)^{l_0}\ra = O(m^{k_0})$ does not necessarily require that
the expectation value receives contributions only from a region of
configuration space of vanishing measure $O(m^{k_0})$ [according to
$D\!A\,\mathcal{P}(m,A)$] in the chiral limit, as assumed above, not
even for positive observables; but it could alternatively be due to
$\mathcal{P}(m,A)$ being supported where $\Oc(A)^{l_0}$ is of
magnitude comparable with $m^{k_0}$, so that
$\chi_{S_m(\mathcal{P})}(A)\Oc(A)^{l_0}=O(m^{k_0})$.  This applies,
e.g., to self-averaging observables, that in the thermodynamic limit
equal their expectation value on the support of $\mathcal{P}(m,A)$.
Under this alternative assumption, $P_{\Oc}(m,A)$ given above
Eq.~\eqref{eq:meas_shr} is not well defined in the chiral limit, so
Eq.~\eqref{eq:aft_crit5} is not guaranteed to hold; in fact, one finds
instead that $\la \Oc(A)^{l}\ra = O(m^{(k_0/l_0)l})$ if
$\la\Oc(A)^{l_0}\ra=O(m^{k_0})$.

It is worth stressing that positivity of $\Oc(A)$ plays no role in the
argument leading to Eq.~\eqref{eq:meas_shr}, and also does not force
one to make the relevant assumption used there concerning the support
of the observable. The alternative possibility just discussed is
equally valid for a positive observable.

\paragraph{}

The discussion above can be made rigorous in the language of
density-of-states functions, rewriting the expectation values
$\la \Oc(A)^l\ra$ as
\begin{equation}
  \label{eq:aft_crit6}
  \la \Oc(A)^l\ra =  \int dx\, x^l p_{\Oc}(x,m)\,,
\end{equation}
where $ p_{\Oc}(x,m) \equiv \left\la \delta(x-\Oc(A))\right\ra$, with
$\int dx\,p_{\Oc}(x,m) = 1$, is the density-of-states function (which
is generally a positive distribution) associated with $\Oc(A)$. For
positive $\Oc(A)$ the support of $p_{\Oc}(x,m)$ is restricted to
$x\ge 0$.  In this language, by claiming that
Eq.~\eqref{eq:aft_crit4_0} implies Eqs.~\eqref{eq:aft_crit4} and
\eqref{eq:aft_crit5} [with $P(m,A)$ having the desired properties],
the authors of Ref.~\cite{Aoki:2012yj} are probably implicitly
assuming that the density-of-states function is of the form
\begin{equation}
  \label{eq:p01}
  p_{\Oc}(x,m) = [1-\alpha(m)]\delta(x) + \alpha(m) \tilde{p}(x,m)\,,
\end{equation}
where $\tilde{p}(x,m)$ is an ordinary positive function supported on
$x\ge 0$, normalized for convenience as a probability distribution,
$\int dx\,\tilde{p}(x,m) =1$, and with moments
$\mu_l(m) \equiv \int dx\,x^l\tilde{p}(x,m)$ bounded in the chiral
limit; and where $0\le \alpha(m)\le 1$ with $\alpha(m)=O(m^{k_0})$. In
this case $\la \Oc(A)^l\ra/ \alpha(m)\to \mu_l(0)$ as $m\to 0$, so
$\la \Oc(A)^l\ra = O(\alpha(m))=O(m^{k_0})$,
$\forall l\ge 1$. Note, however, that finding a positive integer $l_0$
for which $\la \Oc(A)^{l_0}\ra = O(m^{k_0})$ is not sufficient to show
that $\alpha(m)=O(m^{k_0})$, as it is possible that $\mu_{l_0}(0)=0$
and so $\la \Oc(A)^{l_0}\ra$ vanishes in the chiral limit faster than
$\alpha(m)$, and so of those $\la \Oc(A)^{l}\ra$ for which
$\mu_l(0)\neq 0$. This is already a counterexample to the general
statement.

If $p_{\Oc}(x,m)$ is instead a Gaussian restricted to $x\ge 0$,
\begin{equation}
  \label{eq:p02}
  p_{\Oc}(x,m) =  \mathcal{N}(m)^{-1} e^{-\f{(x-x_0(m))^2}{2\sigma(m)^2}}\theta(x)\,, 
\end{equation}
with $\mathcal{N}(m)$ a suitable normalization factor, and if
$x_0(m),\sigma(m)>0$ for $m\neq 0$ but vanish in the chiral limit,
$x_0(0)=\sigma(0)=0$, one finds
\begin{equation}
  \label{eq:aft_crit_6ter}
  \la \Oc(A)^l\ra \mathop \sim_{m\to 0}  \left\{
    \begin{aligned}
      &    C^{(l)}_\varsigma x_0(m)^l\,, & &\text{if } \varsigma\neq \infty \,, \\
      & \tilde{C}^{(l)} \sigma(m)^l\,, & &\text{if }  \varsigma= \infty    \,,
    \end{aligned}\right.
\end{equation}
where $\varsigma \equiv \lim_{m\to 0}\f{\sigma(m)}{x_0(m)}$, and
$C_{\varsigma}^{(l)}$ and $\tilde{C}^{(l)}$ are nonzero constants.
Since $\la \Oc(A)^{l}\ra /\la \Oc(A)^{l_0}\ra \propto x_0(m)^{l-l_0}$
or $\propto \sigma(m)^{l-l_0}$, for $l<l_0$ it blows up as $m\to 0$,
so if $\la \Oc(A)^{l_0}\ra $ vanishes like $m^{k_0}$ in the chiral
limit, then $\la \Oc(A)^{l}\ra$ for $0<l<l_0$ vanishes more slowly and
cannot be $O(m^{k_0})$. This example again contradicts the claim of
Ref.~\cite{Aoki:2012yj}.

An even simpler possibility is of course
$ p_{\Oc}(x,m)=\delta(x-x_0(m))$, with $x_0(m)\ge 0$ vanishing in the
chiral limit. In this case $ \la \Oc(A)^l\ra = x_0(m)^l$, and one
reaches the same conclusions as with the Gaussian functional form.
Since the density-of-states function is a Dirac delta if $\Oc(A)$ is a
self-averaging observable, this counterexample is certainly of
physical relevance.

More generally, if there is some $x_0(m)$, with $x_0(0)=0$, for which
the function $\pi_{\Oc}(x,m)\equiv p_{\Oc}(x/x_0(m),m)$ has a
well-defined chiral limit $\pi_{\Oc}(x,0)$ with finite moments, one
finds
\begin{equation}
  \label{eq:aft_crit6_gen}
  \la \Oc(A)^l\ra
  = x_0(m)^l\int_0^\infty dx\, x^l \pi_{\Oc}(x,0) + o\left(x_0(m)^{l}\right) \,, 
\end{equation}
and one reaches the same conclusions as above.

\paragraph{}

These counterexamples show that without additional assumptions one
cannot deduce Eq.~\eqref{eq:aft_crit5} as a necessary consequence of
Eq.~\eqref{eq:aft_crit4_0}, and more generally one cannot conclude
that $\la \Oc(A)^{l_0}\ra = O(m^{k_0})$ for some $l_0,k_0$ always
implies $\la \Oc(A)^l\ra = O(m^{k_0})$ for any integer $l\ge 1$ for
positive observables that depend only on gauge fields. Of course, one
may assume that the support of the relevant observables is such that
Eqs.~\eqref{eq:aft_crit4} and \eqref{eq:aft_crit5} can be derived [see
the discussion around Eq.~\eqref{eq:meas_shr}], but there seems to be
no compelling reason to do so, and the applicability of such an
assumption in realistic cases is called into question by the
simplicity of the counterexamples.

In Ref.~\cite{Aoki:2012yj} the authors make the additional
$m^2$-analyticity assumption mentioned above, namely that in the
symmetric phase the expectation value in the infinite-volume limit (if
it exists at $m\neq 0$) of an $m$-independent functional of the gauge
fields, $\Oc(A)$, is an analytic function of $m^2$ near $m=0$. It is
not clear whether this assumption can be used to derive
Eq.~\eqref{eq:aft_crit5}, and if so, how; and if it was used in
Ref.~\cite{Aoki:2012yj} to this end. For nonlocal functionals (such as
those used in Ref.~\cite{Aoki:2012yj}) such an assumption cannot be
directly justified using the properties of a local quantum field
theory in the symmetric phase (see Ref.~\cite{Giordano:2025shr}), and
should be regarded as an independent assumption.

In any case, under such an unrestricted $m^2$-analyticity assumption
the counterexamples given above do not apply. In order for
$\la \Oc(A)^l\ra $ with $l$ integer to be analytic in $m^2$ near zero,
$x_0(m)$ must be proportional to $m^{2k}$ for some integer $k$, in all
cases except for the Gaussian density-of-states function if
$\varsigma=\infty$, in which case this requirement applies to
$\sigma(m)$. [Conversely, requiring that $x_0(m)$, and for the
Gaussian density-of-states function also $\sigma(m)$, be analytic
functions of $m^2$ guarantees $m^2$ analyticity of $\la \Oc(A)^l\ra $
for any integer $l$.] It is then clear that $\la\Oc(A)^\alpha\ra$ will
not be $m^2$ analytic near $m=0$ for suitably chosen noninteger
powers $\alpha$. The possibility to derive Eq.~\eqref{eq:aft_crit5}
from Eq.~\eqref{eq:aft_crit4_0} under this additional assumption
remains therefore open.

\paragraph{}

However, an unrestricted $m^2$-analyticity assumption has very
peculiar consequences that lead one to question its applicability. Let
$\Lc(A)$ be any $m$-independent functional that depends only on the
lattice gauge fields $A(y)$ for $y$ in some finite region of the
lattice, kept fixed in the thermodynamic limit. Let $\Lc(A_x)$ be the
translated functionals obtained by replacing $A(y)$ with
$A_x(y)= A(y+x)$ in the argument of $\Lc(A)$, and construct the
``intensive'' gauge observable
\begin{equation}
  \label{eq:an_intens}
  \Oc_{\Lc}(A) \equiv \f{1}{V}\sum_x \Lc(A_x)\,,  
\end{equation}
where the sum runs over the lattice sites and $V$ is the lattice
volume. Since interactions are short range this observable is
self-averaging, and in the thermodynamic limit the corresponding
density-of-states function reads as
\begin{equation}
  \label{eq:an_intens2}
  p_{\Oc_{\Lc}}(x,m) = \delta(x-\ell_{\Lc}(m))\,, \quad
  \ell_{\Lc}(m) \equiv  \la \Oc_{\Lc}(A)\ra\,.
\end{equation}
Here it is assumed that $\ell_{\Lc}(m)$ does not vanish, e.g., due to
symmetry reasons, at $m\neq 0$; note that $\ell_{\Lc}(m)$ must be
analytic in $m^2$ by the $m^2$-analyticity assumption.  An obvious
example of an observable of this type is the usual gauge action
density.  More general examples are the average over lattice
translations, in the sense of Eq.~\eqref{eq:an_intens}, of multilocal
gauge observables, e.g., products of Wilson loops of fixed but
arbitrary shapes and relative positions, which involve gauge fields in
a fixed finite lattice region, although disconnected. Now, for the
observables
$ \tilde{\Oc}_{\Lc}^{(\alpha)}(A) \equiv
|\Oc_{\Lc}(A)-\ell_{\Lc}(0)|^\alpha$, which are again $m$ independent
functionals of gauge fields only, one finds
\begin{equation}
  \label{eq:an_intens4}
  \begin{aligned}
    \la\tilde{\Oc}_{\Lc}^{(\alpha)}(A)\ra
&  = \int dx\, |x-\ell_{\Lc}(0)|^\alpha \,p_{\Oc_{\Lc}}(x,m)\\
&  = |\ell_{\Lc}(m)-\ell_{\Lc}(0)|^\alpha\,,
  \end{aligned}
\end{equation}
that cannot be analytic in $m^2$ near $m=0$ for every $\alpha>0$
unless it vanishes identically for sufficiently small $m$, i.e.,
unless $\ell_{\Lc}(m)= \ell_{\Lc}(0)$ if $|m|<m_{\Lc}$, for some
possibly observable-dependent $m_{\Lc}>0$. [The $m^2$-analyticity
assumption then does not apply for $\alpha<0$, since in this case the
expectation value, Eq.~\eqref{eq:an_intens4}, diverges.] Using
translation invariance, this would imply that
$ \la \Lc( A)\ra = \ell_{\Lc}(m)$ becomes $m$-independent for
$|m|<m_{\Lc}$, and so is not $m^2$ analytic at some finite nonzero
$m=\pm m_{\Lc}$ (unless it were $m$ independent for all $m$, which is
very unlikely; for sure, there exists some $\Lc$ whose expectation
value is not $m$ independent for all $m$).
  
Under the unrestricted $m^2$-analyticity assumption, in the simplest
case where $\inf_{\Lc}m_{\Lc}\neq 0$, there would then be a phase at
small $m$ where all $\la \Lc( A)\ra$ are $m$ independent; otherwise
there would have to be infinitely many phase-transition lines, with
these expectation values becoming $m$ independent gradually. This
would apply in particular to all the $n$-point correlation functions
of local functionals of the gauge fields (in the usual sense of
involving products of gauge link variables in a fixed connected region
of the lattice). The existence of such a phase seems highly unlikely,
even as a lattice artifact, given that it is at small $m$ that the
backreaction of the fermionic determinant on the gauge-field dynamics
is the most effective.

The fact that the unrestricted $m^2$-analyticity assumption implies
the existence of such a peculiar phase at small $m$ calls, if not for
its rejection, at least for some suspicion on its applicability. In
Ref.~\cite{Aoki:2012yj} the $m^2$-analyticity assumption, although
formulated without restrictions, is used explicitly only for certain
nonlocal observables and their integer powers, and so the argument
above would not necessarily invalidate their results (although one
would have to properly justify why $m^2$ analyticity applies to the
relevant observables).  Clearly, if one restricts the
$m^2$-analyticity assumption in this way then the counterexamples
provided above are perfectly valid. (One can of course make the
counterexamples inapplicable to the observables of interest by
requiring $m^2$ analyticity for the expectation value of their
noninteger powers, but such an \textit{ad hoc} assumption would have
to be thoroughly justified.)

\paragraph{}
  
In Sec.~III~G of Ref.~\cite{Aoki:2012yj} [see after their Eq.~(92)],
the implication that if $\la \Oc(A)^{l_0}\ra = O(m^{k_0})$ for some
positive integer $l_0$ then $\la \Oc(A)^l\ra = O(m^{k_0})$ for any
integer $l\ge 1$ (in particular for $1\le l<l_0$) is crucially used,
together with technical assumptions on the spectral density on a fixed
configuration (including in practice a restricted form of the
$m^2$-analyticity assumption), to prove that in the symmetric phase
$\chit$ and $\rho(0;m)$ vanish faster than any power of $m$ in the
chiral limit, and therefore identically below some small but nonzero
$m$ due to analyticity in $m^2$ near $m=0$. However, the
counterexamples discussed above show that the status of this
implication is problematic, and these conclusions of
Ref.~\cite{Aoki:2012yj} should be critically reconsidered. These
counterexamples could be invalidated by making stronger assumptions on
the $m^2$ analyticity of expectation values in the symmetric phase,
but these assumptions would have to be properly justified. In
particular, assuming $m^2$ analyticity for the expectation value of
any $m$-independent gauge-field functional (when this exists at
$m\neq 0$) would lead to a very peculiar phase at small but nonzero
$m$, casting serious doubts on the applicability of this assumption.

\begin{acknowledgments}
  I thank T.~G.~Kov{\'a}cs, G.~Mark\'o, and D.~N\'ogr\'adi for
  discussions, and S.~Aoki and H.~Fukaya for correspondence.  This
  work was partially supported by the NKFIH grants K-147396, NKKP
  Excellence 151482, and TKP2021-NKTA-64.
\end{acknowledgments}

\appendix

\onecolumngrid

\vspace{\stretch{1}}
\newpage

\setcounter{equation}{0}
\setcounter{paragraph}{0}
\setcounter{figure}{0}
\setcounter{table}{0}

\begin{center}
  \rule{300pt}{1pt}

  \textbf{\large Note on the authors' reply to my comment}
  \rule{0pt}{16pt}

\rule{300pt}{1pt}
\end{center}

\vspace{5pt}

{\small\center
  \begin{minipage}{0.785\linewidth}
    In this note I refute the objections raised by S.~Aoki and
    H.~Fukaya in their reply to my comment. I show that both the
    $m^2$-analyticity assumption they invoke against my arguments, and
    the very claim that these arguments criticize, imply that all
    local gluonic correlators are independent of the light-quark mass
    at small mass.  This behavior, however, is not observed in quantum
    chromodynamics.  I then show that their claim of having found a
    mistake in my arguments is baseless.
  \end{minipage}\\}

\vspace{30pt}

\twocolumngrid
\paragraph{}

In their reply~\cite{db_Aoki:2026hzl} to my
comment~\cite{db_Giordano:2025vbb} on a paper of theirs with
Y.~Taniguchi~\cite{Aoki:2012yj}, S.~Aoki and H.~Fukaya (referred to in
the following as ``the authors'' for brevity) dismiss my arguments
against a technical claim of Ref.~\cite{Aoki:2012yj} by claiming that
(i) my counterexamples violate their $m^2$-analyticity assumption, and
(ii) they found a mistake in one of my arguments. Here I refute these
objections and reaffirm the validity of my arguments.

For the benefit of my 2.5 readers I briefly recall the context of the
dispute. The setting is quantum chromodynamics (QCD) in the two-flavor
chiral limit in the symmetric phase, i.e., at temperatures high enough
so that chiral symmetry gets restored as the common mass of the two
lightest quarks, $m$, is sent to zero. The theory is discretized on a
hypercubic lattice with Ginsparg--Wilson fermions for mathematical
definiteness.

The technical claim of Ref.~\cite{Aoki:2012yj} criticized in
Ref.~\cite{db_Giordano:2025vbb} can be summarized as follows: If a
positive, $m$-independent, purely gluonic observable, $\Oc(A)\ge 0$,
satisfies $\la \Oc(A)^{l_0} \ra = O(m^{k_0})$ for some integer
$l_0\ge 1$ and some $k_0>0$, then $\la \Oc(A)^{l} \ra = O(m^{k_0})$
for any integer $l\ge 1$. Here and in the following $\la\ldots\ra$
denotes the expectation value in the thermodynamic limit.  In
Ref.~\cite{Aoki:2012yj} this technical claim is used to show that in
the symmetric phase the topological susceptibility and the spectral
density of the Dirac operator at the origin vanish identically for
sufficiently small masses, i.e., for $|m|<m_0$ for some $m_0>0$.

In Ref.~\cite{db_Giordano:2025vbb} I argued that this technical claim is
generally unfounded and provided counterexamples, concluding that the
validity of its consequences should be reassessed. In
Ref.~\cite{db_Aoki:2026hzl} the authors state that these counterexamples
are not valid, since they lead to violations of the $m^2$-analyticity
assumption on gluonic observables used in
Ref.~\cite{Aoki:2012yj}. However, in Ref.~\cite{db_Giordano:2025vbb} I
also showed that this assumption, taken without restrictions on the
observables (beyond minimal ones to make it workable), leads to
peculiar and physically unlikely consequences for local gluonic
correlators, which strongly suggests that this form of the assumption
is unjustified and cannot be used to invalidate my counterexamples.
In Ref.~\cite{db_Aoki:2026hzl} the authors claim to have found a mistake
in this argument.

\paragraph{}

The issue with objection (i) is that it ignores the main problem of
the $m^2$-analyticity assumption as formulated by the authors, namely
the fact that its peculiar consequences are not supported by any
actual evidence.

In Ref.~\cite{db_Aoki:2026hzl} the authors state that the
$m^2$-analyticity assumption is ``a crucial assumption of QCD at high
temperatures'' (Sec.~IV, \S1). Rather than an assumption, in the sense
of axiom, \textit{of} QCD at high temperature, the $m^2$-analyticity
assumption is a conjecture \textit{on} the analyticity properties of
this theory.  In principle, these properties could be derived from the
fundamental axioms, and it is only our inability to do so that forces
us to conjecture what they are; of course, our conjectures could be
wrong.  The $m^2$-analyticity assumption may then be crucial for
certain studies of high-temperature QCD, but if it turned out to be
wrong the theory itself would be unaffected.

To invoke the $m^2$-analyticity assumption against the counterexamples
of Ref.~\cite{db_Giordano:2025vbb} one needs first of all to define
precisely what this assumption, or rather conjecture, is, and then
check that its consequences are not in contradiction with the actual
properties of QCD. In Ref.~\cite{db_Aoki:2026hzl} the authors formulate
the $m^2$-analyticity assumption as follows: ``every gluonic
observable in the chirally symmetric phase is an analytic function of
$m^2$'' (Sec.~I, \S1). ``Observable'' should be understood here as the
expectation value of an observable, i.e., of a functional of the
fields of the theory (see Ref.~\cite{Aoki:2012yj}, Sec.~II~C). In this
note I use ``observable'' exclusively to denote these functionals.
When the authors write ``every'' they surely do not mean it literally
(as it is clear from the discussion in Ref.~\cite{Aoki:2012yj},
Sec.~II~C). An arbitrary explicit dependence of the observable on $m$
is certainly not allowed; and since analyticity assumptions on the $m$
dependence are meaningful only in the thermodynamic limit, the
expectation value of the relevant observables must be non-divergent in
this limit. The ``reasonable'' observables to which the
$m^2$-analyticity assumption may apply are then $m$-independent
functionals of the gauge fields with non-divergent expectation value
in the thermodynamic limit.  These include products of local
observables and their suitably normalized spacetime integrals, but
also genuinely nonlocal functionals of the gauge fields (e.g., the
square root of the plaquette).

As in Ref.~\cite{db_Giordano:2025vbb}, the assumption that the
expectation value in the thermodynamic limit of every reasonable
gluonic observable is an analytic function of $m^2$ near $m=0$ will be
referred to as the ``unrestricted $m^2$-analyticity assumption''. The
assumption of $m^2$ analyticity is well motivated for correlators of
local gluonic observables and their integrals, as they are at least
$C^\infty$ functions of $m^2$ in the symmetric phase if the
correlation length does not diverge in the chiral limit (see
Ref.~\cite{Giordano:2025shr}, Sec.~III). For nonlocal gluonic
observables, on the other hand, this assumption is debatable and
should be separately justified (for spectral observables see
Ref.~\cite{Giordano:2025shr}, Sec.~III). These observables must be
included in the assumption if the authors want to invoke it to
invalidate my counterexamples.

\paragraph{} 

In Ref.~\cite{db_Giordano:2025vbb} I showed that under the unrestricted
$m^2$-analyticity assumption the expectation value in the
thermodynamic limit of an arbitrary multilocal gluonic observable must
become exactly $m$ independent for small $m$. Multilocal gluonic
observables are $m$-independent gauge-invariant polynomials of the
lattice gauge fields (i.e., of the link variables) in a finite,
possibly disconnected region of the lattice, kept fixed as the size of
the lattice grows; examples are the products of any number of Wilson
loops of fixed but arbitrary shapes, sizes, and positions.  The
expectation values in the thermodynamic limit of multilocal gluonic
observables will be called ``local gluonic correlators'' for brevity.
 
The argument of Ref.~\cite{db_Giordano:2025vbb} makes use of observables
of the form
\begin{equation}
  \label{eq:1}
  \Oc_{\Lc}(A) \eqdef \f{1}{V}  \sum_x \Lc(A_x)\,,
\end{equation}
where $\Lc(A_x)$ is obtained from a multilocal gluonic observable,
$\Lc(A)$, by translating the gauge fields by the lattice vector $x$,
$V$ is the lattice volume, and the sum is over all lattice sites. For
systems with short-range interactions the observables $\Oc_{\Lc}(A)$
are self-averaging, so in the thermodynamic limit
$\la \Oc_{\Lc}(A)^n \ra = \ell_{\Lc}(m)^n$, where
$\ell_{\Lc}(m) \eqdef \la \Lc(A)\ra $ and I used translation
invariance, and the density-of-states function,
$ p_{\Oc}(x;m) \eqdef\la \delta(x - \Oc(A))\ra$, is a Dirac delta,
i.e.,\footnote{This applies also if $\ell_{\Lc}(m)$ vanishes, so the
  assumption of Ref.~\cite{db_Giordano:2025vbb} that
  $\ell_{\Lc}(m)\neq 0$ for $m\neq 0$ is unnecessary and can be
  dropped.}
\begin{equation}
  \label{eq:loc}
  p_{\Oc_{\Lc}}(x;m) = \delta(x- \ell_{\Lc}(m))\,.
\end{equation}
As a consequence, $\la f(\Oc_{\Lc}(A)) \ra = f(\ell_{\Lc}(m))$ for any
continuous function $f$. Taking
$f_\alpha(x) \eqdef |x-\ell_{\Lc}(0)|^\alpha$, it is easy to see that
there is no function $\ell_{\Lc}(m)$, analytic in $m^2$ at $m=0$, such
that $f_\alpha(\ell_{\Lc}(m))$ is analytic in $m^2$ at $m=0$ for every
$\alpha\in\mathbb{R}^+$, other than functions identically constant
near zero.  It follows that under the unrestricted $m^2$-analyticity
assumption all local gluonic correlators must be $m$ independent, at
least for $|m|$ below some (possibly correlator-dependent) nonzero
value. More generally, the expectation value in the thermodynamic
limit of every observable $\Sc(A)$ for which
$p_{\mathcal{S}}(x;m)=\delta(x-\la \Sc(A)\ra)$ holds must become $m$
independent at small $m$.

By a similar reasoning, one can show directly that the technical claim
of Ref.~\cite{Aoki:2012yj} cannot be correct unless every local
gluonic correlator, and more generally every $\la\Sc(A)\ra$, becomes
$m$ independent at small $m$, independently of any analyticity
assumption. In fact, for the positive $m$-independent gluonic
observable $\Pc_{\Lc}(A) \eqdef (\Oc_{\Lc}(A)-\ell_{\Lc}(0))^2$ one has
in the thermodynamic limit
$\la\Pc_{\Lc}(A)^n\ra=\dl_{\Lc}(m)^{2n}=o(m^0)$, where
$\dl_{\Lc}(m) \eqdef\ell_{\Lc}(m)-\ell_{\Lc}(0)$. If $\dl_{\Lc}(m)$ is
not identically zero at small $m$ one finds that, e.g.,
$\la\Pc_{\Lc}(A)^2\ra$ is $O(\dl_{\Lc}(m)^2)$ but $\la\Pc_{\Lc}(A)\ra$
is not, contradicting the technical claim.

Complete $m$ independence of all local gluonic correlators is
certainly excluded in QCD, so if the unrestricted $m^2$-analyticity
assumption or the technical claim were correct these correlators would
have to become $m$ independent below some nonzero value of $m$; here
they would be non-analytic, thus signaling a phase transition.  The
predicted $m$ independence at small $m$ is suspicious from the
theoretical point of view, even as a lattice artifact, as it is
unexpected when the back-reaction of the fermion determinant on the
gauge fields should be the most effective. Rather than insisting on
the unrestricted assumption and having to accept this peculiar
behavior, it is simpler (and theoretically harmless) to restrict the
assumption, and accept that not all (reasonable) nonlocal gluonic
observables have an expectation value in the thermodynamic limit that
is $m^2$ analytic at zero; and similarly give up the general validity
of the technical claim. More importantly, in QCD toward the chiral
limit there is no evidence, numerical or otherwise, for this peculiar
behavior. The unrestricted $m^2$-analyticity assumption is then a
conjecture with no good theoretical basis and no supporting evidence,
and is most likely wrong in QCD. Similarly, the technical claim of
Ref.~\cite{Aoki:2012yj} is most likely invalid in the general case.

Clearly, one cannot invoke conjectures with no support as a
counterargument, or any debate would become impossible.  Concerning
objection (i), the counterexamples of Ref.~\cite{db_Giordano:2025vbb}
cannot then be dismissed by invoking the unrestricted
$m^2$-analyticity assumption, and remain valid. More directly, and
independently of any analyticity assumption, the existence of a single
local gluonic correlator not identically constant at small $m$
suffices to explicitly disprove the technical claim of
Ref.~\cite{Aoki:2012yj}.

\paragraph{}

The authors of Ref.~\cite{db_Aoki:2026hzl} dismiss the validity of the
predictions obtained under the unrestricted $m^2$-analyticity
assumption in Ref.~\cite{db_Giordano:2025vbb}, stating that
Eq.~\eqref{eq:loc} is generally incorrect (Sec.~III). Here the authors
are attacking a straw man: only observables $\Oc_{\Lc}(A)$ obtained
via Eq.~\eqref{eq:1} from some multilocal $\Lc(A)$ are used in the
argument of Ref.~\cite{db_Giordano:2025vbb} discussed above, and
Eq.~\eqref{eq:loc} holds for observables of this type.  The fact that
Eq.~\eqref{eq:loc} does not hold for a generic observable is
completely irrelevant, and objection (ii) is baseless.

The authors propose the following counterexample,
\begin{equation}
  \label{eq:nloc}
  \Nc(A_x) \eqdef \sum_y q(y) q(x) = Q q(x)\,, 
\end{equation}
with $q(x)$ the topological charge density and $Q$ the topological
charge. This is not a multilocal observable, as it involves gauge
fields across the whole lattice, and intrinsically depends on the
system size, so Eq.~\eqref{eq:loc} does not have to apply to
$\Oc_{\Nc}(A)= \sum_x \Nc(A_x)/V= Q^2/V$. In no way does then
Eq.~\eqref{eq:nloc} constitute a valid counterexample to
Eq.~\eqref{eq:loc}, which holds for $\Oc_{\Lc}(A)$ obtained from a
multilocal $\Lc(A)$, but is not guaranteed to hold otherwise.

The authors conclude that ``[t]he ``peculiar'' consequences obtained
by the author of [\cite{db_Giordano:2025vbb}] do not arise from the
assumption of $m^2$-analyticity, but rather from the incorrect
assumption [\eqref{eq:loc}] about the density of states'' (Sec.~III).
The validity of Eq.~\eqref{eq:loc} for quantities of the form
Eq.~\eqref{eq:1} is not assumed when $\Lc(A)$ is multilocal, but
proved as a consequence of the short-range nature of interactions in
the system under consideration. The peculiar consequences found in
Ref.~\cite{db_Giordano:2025vbb} and above do indeed originate in the
unrestricted $m^2$-analyticity assumption, and not in the fact that
Eq.~\eqref{eq:loc} does not hold for observables for which it has no
reason to hold.

\paragraph{}

Besides the two objections just refuted, in Ref.~\cite{db_Aoki:2026hzl}
there are a few more statements that I would like to address.

In Sec.~I, \S1 the authors base their $m^2$-analyticity assumption on
the $m^2$ analyticity of the partition function above the critical
temperature; and in \S4 they propose the partition function as an
example of nonlocal object for which $m^2$ analyticity at $m=0$ is
justified. The partition function, however, is not an observable in
any usual sense, and is not even well defined in the thermodynamic
limit, where the $m^2$-analyticity assumption should be formulated.
On the other hand, in any finite volume it is $m^2$ analytic also in
the spontaneously broken phase. Perhaps what the authors have in mind
is the free energy density, for which assuming $m^2$ analyticity in
the symmetric phase is meaningful and acceptable; this would still not
justify making this assumption for arbitrary (reasonable) nonlocal
observables.

In Sec.~II the authors discuss the possibility of writing the
expectation value of an $m$-independent gluonic observable, $\Oc(A)$,
satisfying $\la \Oc(A)^{l_0} \ra = o(m^k)$ for some integer $l_0\ge 1$
and some $k\ge 0$, in the form
\begin{equation}
  \label{eq:7}
  \la \Oc(A)^{l_0} \ra = m^{k_0} \int DA\,P(m,A)\Oc(A)^{l_0}=O(m^{k_0}) \,,
\end{equation}
for some $k_0>k$ and some $P(m,A)$, as done in
Ref.~\cite{Aoki:2012yj}.  In this discussion they focus on the
well-definedness of $P(m,A)$ in a finite volume at nonzero mass, which
is obvious and was never in question; and in the thermodynamic limit
at nonzero mass, which can certainly be accepted, at least at the
formal level. What they carefully avoid is discussing the existence of
$P(0,A)$ in the subsequent chiral limit, which is far from trivial,
and the crux of the matter for what concerns proving the technical
claim of Ref.~\cite{Aoki:2012yj}. The authors state that the
\textit{integral} in Eq.~\eqref{eq:7} ``should be an analytic function
which satisfies all the required assumptions'' (Sec.~II, \S3), but
this does not imply in any way the existence of a well-defined chiral
limit of the \textit{integrand}.

In Sec.~II, \S2 the authors write that they believe that
$\la \Oc(A)^{l_0} \ra = o(m^k)$ is ``almost equivalent'' with
Eq.~\eqref{eq:7}, and that in Ref.~\cite{db_Giordano:2025vbb} I ``admit[]
that the equivalence is not crucial for the main discussion.''  I
certainly did not admit such a thing, since Eq.~\eqref{eq:7} is
clearly the first step of Ref.~\cite{Aoki:2012yj} towards proving the
technical claim. In Ref.~\cite{db_Giordano:2025vbb}, though, I made it
clear that this equivalence is useless for this purpose unless it
applies to the chiral limit as well, i.e., unless one can argue that a
well-defined $P(0,A)$ exists. Affirming the equivalence of the two
expressions without addressing this issue is misleading at best.

\paragraph{}

In conclusion, the unrestricted $m^2$-analyticity assumption, if
correct, would imply that all local gluonic correlators become $m$
independent at sufficiently small $m$. Independently of this
assumption, the same behavior would follow if the technical claim of
Ref.~\cite{Aoki:2012yj} were correct. The claim of
Ref.~\cite{db_Aoki:2026hzl} of having found a mistake in the argument of
Ref.~\cite{db_Giordano:2025vbb} leading to these results is based on a
straw man argument, and so it is actually baseless.

The behavior predicted by the unrestricted $m^2$-an\-a\-lyt\-ic\-i\-ty
assumption, and independently by the technical claim of
Ref.~\cite{Aoki:2012yj}, is theoretically suspicious and has no
supporting evidence in QCD. This makes both the applicability of this
assumption to QCD and the general validity of this technical claim
highly questionable. It is theoretically much more reasonable to
accept restrictions on the $m^2$-analyticity assumption for what
concerns genuinely nonlocal observables; and that the technical claim
of Ref.~\cite{Aoki:2012yj} is not valid in the general case. This does
not conflict with any fundamental principle, and with the currently
available evidence.  However, the prediction of
Ref.~\cite{Aoki:2012yj} that the topological susceptibility and the
spectral density of the Dirac operator at the origin vanish
identically at small $m$ in the symmetric phase of QCD, based on the
aforementioned technical claim, is left without any theoretical
backing.

\mbox{}

\begin{acknowledgments}
  I thank T.~G.~Kov{\'a}cs, G.~Mark\'o, and D.~N\'ogr\'adi for
  discussions. This work was partially supported by the NKFIH grants
  K-147396, NKKP Excellence 151482, and TKP2021-NKTA-64.
\end{acknowledgments}

\bibliographystyle{../../../../apsrev4-2_mod}
\bibliography{../../../../references_chi_PRD}

\end{document}